\documentclass[12pt,a4paper]{article}
\usepackage[top=25truemm,bottom=35truemm,left=25truemm,right=25truemm]{geometry}
\usepackage{amsmath,amssymb}
\usepackage{graphicx}

\begin{document}
\setlength{\baselineskip}{18pt}
\begin{titlepage}
\begin{flushright}
\begin{tabular}{l}
 SU-HET-12-2014\\
\end{tabular} 
\end{flushright}

\vspace*{1.2cm}
\begin{center}
{\Large\bf Hierarchy problem, gauge coupling unification at the Planck scale,
 and vacuum stability

}
\end{center}
\lineskip .75em
\vskip 1.5cm

\begin{center}
{\large Naoyuki Haba}$^1$,
{\large Hiroyuki Ishida}$^1$,
{\large Ryo Takahashi}$^1$, and
{\large Yuya Yamaguchi}$^{1,2}$\\

\vspace{1cm}

$^1${\it Graduate School of Science and Engineering, Shimane University,\\
 Matsue 690-8504, Japan}\\
$^2${\it Department of Physics, Faculty of Science, Hokkaido University,\\
 Sapporo 060-0810, Japan}\\

\vspace{10mm}
{\bf Abstract}\\[5mm]
{\parbox{13cm}{\hspace{5mm}
From the point of view of the gauge hierarchy problem,
 introducing an intermediate scale in addition to TeV scale and the Planck scale
 ($M_{\rm Pl} = 2.4 \times 10^{18}\,{\rm GeV}$) is unfavorable.
In that way, a gauge coupling unification (GCU) is expected to be realized at $M_{\rm Pl}$.
We explore possibilities of GCU at $M_{\rm Pl}$ by adding a few extra particles with TeV scale mass
 into the standard model (SM).
When extra particles are fermions and scalars (only fermions) with the same mass,
 the GCU at $M_{\rm Pl}$ can (not) be realized.
On the other hand, when extra fermions have different masses,
 the GCU can be realized around $\sqrt{8 \pi} M_{\rm Pl}$ without extra scalars.
This simple SM extension has two advantages
 that a vacuum becomes stable up to $M_{\rm Pl}$ ($\sqrt{8 \pi} M_{\rm Pl}$)
 and a proton lifetime becomes much longer than an experimental bound. 
}}
\end{center}
\end{titlepage}

\section{Introduction}
The collider experiments have discovered all particles in the standard model (SM),
 and properties of the SM particles are gradually revealed.
Especially, masses of the Higgs boson and top quark are important
 to investigate a behavior of the quartic coupling of the Higgs boson at a high energy scale.
The measurement of Higgs mass showed $125.6 \pm 0.35\,{\rm GeV}$ \cite{Hahn:2014qla},
 and a recent combine analysis of the collider experiments reported the top mass as
 $173.34 \pm 0.76\,{\rm GeV}$ \cite{ATLAS:2014wva}.
A running of the quartic coupling of the Higgs becomes to the negative around $10^{10}\,{\rm GeV}$ GeV
 by use of  the experimental values of the Higgs and top masses.
This behavior seems to indicate that our vacuum is metastable.

There are several ways to make the vacuum stable.
A simple way is to add an extra scalar to the SM.
When we assign odd parity to it under an extra $Z_2$ symmetry,
 it can be a dark matter \cite{Silveira:1985rk}-\cite{Boucenna:2014uma}.
Another way to stabilize the vacuum is modifying runnings of the gauge coupling constants.
It decreases (increases) the values of the top Yukawa (Higgs self-) coupling at a high energy,
 where the vacuum becomes stabilize.
In this paper, we try to realize the gauge coupling unification (GCU) at the Planck scale
 by introducing additional particles in the TeV scale.
This extension really induces the above modification of runnings of the gauge coupling constants.

The so-called hierarchy problem is related to the Higgs sector in the SM.
A quadratic divergence of the Higgs mass seems to be a dangerous problem.
However, the Bardeenfs argument \cite{Bardeen:1995kv} says that it is an unphysical
 because it can be removed by a subtractive renormalization.\footnote{
Reference\,\cite{Tavares:2013dga} pointed out the Bardeen's argument is incorrect,
 and then discussions of the GCU is changed from ours \cite{Giudice:2014tma}.
However, their conclusions completely depend on the way to deal with gravity.
Thus, we do not care about their considerations in this paper.
}
Once it is subtracted and the Higgs mass term is vanishing at the UV scale,
 it continues to be zero toward the lower energy scale,
 since the renormalization group equation (RGE) of the Higgs mass term is proportional to itself.
We assume a classical conformal symmetry to justify the vanishing Higgs mass term at the high energy scale.
This symmetry can be radiatively broken by Coleman-Weinberg mechanism \cite{Coleman:1973jx}.
We can see this situation, for example, in a model with an additional $U(1)$ gauge symmetry
 and three right-handed neutrinos \cite{Iso:2009ss}.
Note that the right-handed neutrinos do not change the running of the SM gauge couplings up to the one-loop level,
 so they are not useful to realize the GCU at the Planck scale.

On the other hand, a logarithmic divergence remains a physical quantity after the renormalization.
When there is a heavy particle with the mass, $M$, which couples the Higgs doublet,
 a quantum correction of $M^2 \log (\Lambda /\mu)$ causes the hierarchy problem.
Thus, naively, we should not introduce any intermediate scales between TeV and UV scales.
We assume here that the UV scale is the Planck scale,
 where all quantum corrections to the Higgs mass are completely vanishing.
This  assumption requires that corrections from breaking effects of the grand unification at the Planck scale
 are canceled by a boundary condition of the UV complete theory.
Although this assumption seems to be artificial, some UV complete theories, e.g., the string theory,
 really provides such a boundary condition.

In addition to the above discussion about the hierarchy problem (for example, Ref.\,\cite{Iso:2012jn}),
 we mention gravity, which involves a specific scale, i.e., the Planck scale.
In the point of view of the classical conformal symmetry,
 there should be no specific scales and no higher-dimensional operators at the classical level.
Thus, a certain scale including the Planck scale should be generated by some dynamics.
For this purpose,
 it is known that the Planck scale arises from the vacuum expectation value of a SM gauge singlet scalar,
 which has a non-minimal coupling with the curvature \cite{Zee:1978wi,Foot:2007iy}.
Since a mechanism of generating the VEV depends on the hidden sector,
 the situation is the same as the above discussion in the decoupling limit between the singlet scalar and the Higgs.
Then, the hierarchy problem can be solved by a boundary condition at the Planck scale,
 in which the Higgs mass term is completely vanishing.

For contributions of gravity to the gauge couplings, they could not be ignored around the Planck scale.
Then they might upset discussion of the GCU at the Planck scale.
To solve this problem,
 it is known that the GCU could be realized due to the asymptotic safety of gravity,
 in which all gauge couplings rapidly become zero and approach the same value around the Planck scale.
In this scenario, the gravitational contributions have been calculated
 at lowest nontrivial order in perturbation theory \cite{Robinson:2005fj}.
However, it is pointed out that
 this calculation depends on a regularization scheme and/or a choice of gauge fixing \cite{Pietrykowski:2006xy}.
In addition, if one applies the dimensional regularization for the calculation,
 there are no gravitational corrections for the gauge couplings.
Thus, we do not consider the gravitational corrections in this paper.

In this paper, we will consider that the Planck scale is the bound of the UV complete theory,
 in which we assume corrections of the Higgs mass term are completely vanishing at the scale.
We also assume that the Higgs mass term is generated by Coleman-Weinberg mechanism
 and it does not cause the hierarchy problem.
In this background, we will consider the GCU at the Planck scale
 to avoid the introduction of any intermediate scales except for the TeV scale.
We introduce extra particles with masses around the TeV scale.
In order to avoid the gauge anomaly, the additional fermionic particles are introduced as vector-like.
A naive analysis will show that,
 when all extra particles are fermions and their masses are the same,
 the GCU at the Planck scale cannot be realized.
On the other hand, when extra particles include some scalars,
 the GCU at the Planck scale can be realized.
Then, we find that there are a number of models which can realize the GCU at the Planck scale.
Next, we will consider another situation, in which extra fermions have different masses.
In this case, models with only extra fermions (no scalars) can realize the GCU around $\sqrt{8\pi}M_{\rm Pl}$.
These extensions make the gauge couplings strong enough to realize the GCU,
 and the top Yukawa (Higgs self-) coupling becomes smaller (larger) than that of the SM at a high energy scale.
Then, the vacuum becomes stable.

This paper is composed as follows.
At first, we will give a brief review of the vacuum stability and related researches in the SM
 in Sec.\,\ref{sec:vacuum}.
Next, we will investigate possibilities for the realization of GCU at some high energy scales
 in Sec.\,\ref{sec:requirement},
 and show conditions of the GCU at the Planck scale
 in Sec.\,\ref{sec:condition}.
Then, examples of extra particles, which satisfy the conditions, are given
 in Sec.\,\ref {sec:realization}.
In addition, we will consider other possibilities,
 in which the GCU can be realized only by extra fermions,
 in Sec.\,\ref {sec:other}.
Finally, summary and discussion are given
 in Sec.\,\ref {sec:summary}.

\section{The vacuum stability} \label{sec:vacuum}
We give a brief review of the vacuum stability and related researches in the SM.
Realization of the vacuum stability depends on a value of the Higgs quartic coupling $\lambda$.
A running of $\lambda$ is obtained by solving the RGE
 $d\lambda/d\ln \mu = \beta_\lambda$, in which $\mu$ is a renormalization scale
 and $\beta_\lambda$ is the $\beta$-function of $\lambda$.
The $\beta$-function of $\lambda$ up to two-loop level is given by \cite{Ford:1992pn,Buttazzo:2013uya}
\begin{eqnarray}
	\beta_{\lambda} &=& \frac{1}{(4\pi)^2} \left[ \lambda \left( 24 \lambda + 12 y_t^2 - \frac{9}{5}g_1^2 - 9 g_2^2 \right)
		- 6 y_t^4 + \frac{27}{200}g_1^4 + \frac{9}{8}g_2^4 + \frac{9}{20}g_1^2 g_2^2 \right] \nonumber\\
		&& + \frac{1}{(4\pi)^4} \left[ \lambda^2 \left( - 312 \lambda - 144 y_t^2 +\frac{108}{5}g_1^2 + 108 g_2^2\right)
		+ \lambda y_t^2 \left( - 3 y_t^2 + \frac{17}{2}g_1^2  \right. \right. \nonumber\\
		&& \left. + \frac{45}{2}g_2^2 + 80 g_3^2 \right) + \lambda \left( \frac{1887}{200}g_1^4 - \frac{73}{8}g_2^4 + \frac{117}{20}g_1^2 g_2^2 \right)
		+ y_t^4 \left( 30 y_t^2 - \frac{8}{5}g_1^2 - 32 g_3^2 \right) \nonumber\\
		&& \left. + y_t^2 \left( - \frac{171}{100}g_1^4 - \frac{9}{4}g_2^4 + \frac{63}{10} g_1^2 g_2^2 \right) 
		- \frac{3411}{2000}g_1^6 + \frac{305}{16}g_2^6 - \frac{1677}{400}g_1^4 g_2^2 - \frac{289}{80}g_1^2 g_2^4 \right],
\end{eqnarray}
 where the top Yukawa and the gauge couplings are included.
Other Yukawa couplings are omitted,
 since they are small enough to be neglected.
For the Higgs pole mass of $M_h = 125.7$\,GeV and the top pole mass of $M_t = 173.3$\,GeV,
 $\lambda$ becomes negative at $\mu \simeq 10^{10}$\,GeV
 and the value of $\lambda$ remains negative up to the Planck scale in the SM.
As a result, the electroweak (EW) vacuum becomes meta-stable.
Thus, one should extend the SM at $\mu \lesssim 10^{10}$\,GeV
 in order to make the vacuum stable with the current center values of Higgs and top masses.

\begin{figure}[t]
  \begin{center}
          \includegraphics[clip, scale=1]{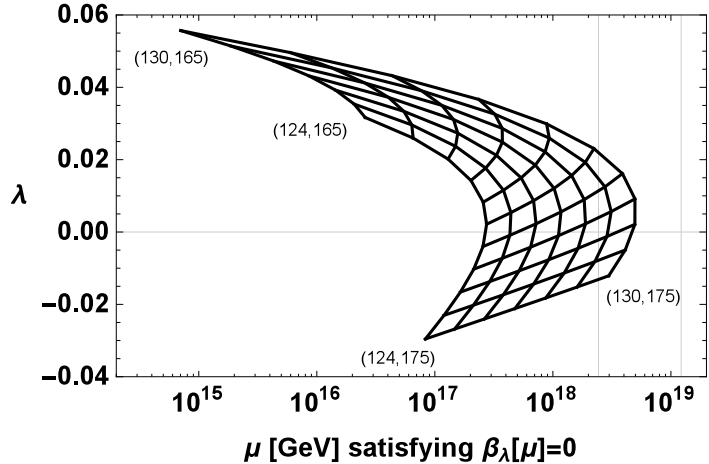}
  \end{center}
\caption{Relation between the energy, where $\beta_\lambda (\mu)=0$ is satisfied,
 and the value of $\lambda$.
The values in parentheses indicate the Higgs and the top pole masses ($M_h$, $M_t$),
 and each width of contours in the lattice corresponds to a change of 1\,GeV for $M_h$ and $M_t$.
Two vertical grid lines represent $M_{\rm Pl}$ and $\sqrt{8\pi}M_{\rm Pl}$, respectively.}
\label{minimum}
\end{figure}

In Fig.\,\ref{minimum},
 we show a relation between the energy, where $\beta_\lambda (\mu)=0$ is satisfied,
 and its value of $\lambda$.
For ($M_h$, $M_t$)=(125.7\,GeV, 173.3\,GeV),
 $\lambda$ is minimized at $\mu \simeq 4.0\times 10^{17}$\,GeV,
 and the value is about $-0.0136$.
If the minimal value of $\lambda$ is zero with $\beta_\lambda =0$ at some high energy scales,
 the vacua at the EW and the high energy scale are degenerate.
This requirement is known as the multiple point criticality principle (MPCP) \cite{Froggatt:1995rt}.
Note that the MPCP can be realized at ${\cal O} (10^{17})$\,GeV
 by use of a lighter top mass as 171\,GeV
 (see also Refs.~\cite{Buttazzo:2013uya} and \cite{Holthausen:2011aa}-\cite{Haba:2014qca}
 for more recent analyses).

From Fig.\,\ref{minimum}, we can show a minimum of the Higgs potential.
It is given by $V_{\rm eff}(\phi) = \frac{1}{4}\lambda \phi^4$,
 where $\phi$ is a field value of the Higgs,
 and its stationary condition satisfies $\beta_\lambda+4\lambda=0$.
This equation is satisfied when $|\lambda|$ becomes almost zero,
 and its solutions are classified in three cases as follows.
\begin{itemize}
\item
$\lambda=0$ and $\beta_\lambda=0$:
 this is just the MPCP condition, 
 where the height of the potential becomes zero.

\item
$\lambda>0$ and $\beta_\lambda<0$:
 this point is a local maximum before $\lambda$ becomes a minimal value.
If there is another solution for $\lambda>0$ and $\beta_\lambda<0$,
 the point is a local minimum.

\item
$\lambda<0$ and $\beta_\lambda>0$:
 this point is a global minimum.
\end{itemize}
For ($M_h$, $M_t$)=(125.7\,GeV, 173.3\,GeV),
 the Higgs potential has a local maximum and global minimum
 at $\phi \simeq 9.5 \times 10^9$\,GeV and $\phi \simeq 3.9 \times 10^{29}$\,GeV, respectively.\footnote{
In this paper, the strong coupling is taken by $\alpha_3 (M_Z) = 0.1184$.}
When $M_h$ is larger than 125.7\,GeV
 and/or $M_t$ is smaller than 173.3\,GeV,
 the points of local maximum and global minimum are larger and smaller, respectively.
For $M_t \lesssim 171.2043$\,GeV,
 the potential is positive in any energy scale,
 and there are no global minimum in the high energy scale.
Only for 171.2041\,GeV $\lesssim M_t \lesssim 171.2043$\,GeV,
 the potential has a local minimum
 at $4.7\times 10^{17}$\,GeV $\lesssim \phi \lesssim 6.1\times 10^{17}$\,GeV.
When the potential has a plateau around the local minimum,
 the Higgs inflation can be realized.
However, if the Higgs potential includes new contributions as higher order terms of $\phi$,
 they can significantly affects the vacuum stability \cite{Branchina:2013jra}-\cite{Branchina:2014usa}.

\section{Requirement for the GCU} \label{sec:requirement}
In this section, we investigate possibilities for the realization of GCU at some high energy scales.
In order to see the behavior of the gauge couplings in an arbitrary high energy scale
 we have to solve the corresponding RGEs.
The one-loop level RGEs of the gauge couplings $\alpha_i=g_i^2/4\pi$ are given by
\begin{eqnarray}
	\frac{d\alpha_i^{-1}}{d\ln \mu} = -\frac{b_i}{2\pi},
\label{RGE}
\end{eqnarray}
 where $i=Y$, 2, and 3,
 and the coefficients of $U(1)_Y$, $SU(2)_L$, and $SU(3)_C$ gauge couplings are given by
 ($b_Y^{\rm SM}$, $b_2^{\rm SM}$, $b_3^{\rm SM}$)=(41/6, $-19/6$, $-7$) in the SM.
$b_1^{\rm SM}$ is obtained by multiplying a GUT normalization factor 3/5 to $b_Y^{\rm SM}$ as $b_1^{\rm SM} = 41/10$.\footnote{
Although the normalization factor of hypercharge depends on GUT models,
 for simplicity, we only consider the factor is 3/5 as in $SU(5)$ GUT.}
Once particle contents in the model are fixed,
 values of $b_i$ are calculated by \cite{Jones:1981we}
\begin{eqnarray}
	b_i = \left[ -\frac{11}{3}c^i_1 + \frac{2}{3} \kappa \sum_{R_f}c_2(R_f) \prod_{j\neq i} d_j(R_f)
		+\frac{1}{3} \eta \sum_{R_s}c_2(R_s) \prod_{j\neq i} d_j(R_s) \right],
\label{b}
\end{eqnarray}
 where $j=Y$, 2, and 3.
The meanings of the notation are as follows:
\begin{itemize}
\item $R_f$, $R_s$: irreducible chiral fermion and scalar representations, respectively
\item $d_i(R)$: dimension of the representation $R$ under the gauge groups
\item $c_2(R)$: quadratic Casimir operator of the representation $R$
\item $c^i_1$: constant usually taken as $c^i_1 = c_2(R^{adj})$ ($c^i_1=N$ for $SU(N)$, and 0 for $U(1)$)
\end{itemize}
Some values of $c_2(R)$ are given in Table \ref{c2} in a convention \cite{Slansky:1981yr, Lindner:1996tf}.
The factor $\kappa$ is 1 or 1/2 for Dirac or Weyl fermions, respectively.
In addition, the factor $\eta$ is 1 or 1/2 for complex or real scalars, respectively.
Using the values, we can obtain contributions to $b_i$ from fermions and scalars.

\begin{table}[t]
\begin{center}
\begin{tabular}{|cc|cc|}\hline
Representation of $SU(2)$ & $c_2$ & Representation of $SU(3)$ & $c_2$ \\
\hline \hline
2 & 1/2  & 3 & 1/2 \\
3 & 2 & 6 & 5/2 \\
4 & 5 & 8 & 3 \\
5 & 10 & 10 & 15/2 \\
\hline
\end{tabular}
\caption{$c_2(R)$ for irreducible representations of $SU(2)$ (left) and $SU(3)$ (right).}
\label{c2}
\end{center}
\end{table}

\begin{table}[t]
\begin{center}
\begin{tabular}{|c|c|}\hline
Irreducible representation & Contribution to ($b_1$, $b_2$, $b_3$) \\
($SU(3)_C$, $SU(2)_L$, $U(1)_Y$) & by fermions\\
\hline \hline
(1, 1, 0) & (0, 0, 0)\\ \hline
(1, 1, $a$)$\oplus$(1, 1, $-a$) & ($\frac{1}{5}a^2$, 0, 0)\\ \hline
(1, 2, $a$)$\oplus$(1, 2, $-a$) & ($\frac{2}{5}a^2$, $\frac{2}{3}$, 0)\\ \hline
(1, 3, 0) & (0, $\frac{4}{3}$, 0)\\ \hline
(1, 3, $a$)$\oplus$(1, 3, $-a$) & ($\frac{3}{5}a^2$, $\frac{8}{3}$, 0)\\ \hline
(3, 1, $a$)$\oplus$($\overline{3}$, 1, $-a$) & ($\frac{3}{5}a^2$, 0, $\frac{2}{3}$)\\ \hline
(3, 2, $a$)$\oplus$($\overline{3}$, 2, $-a$) & ($\frac{6}{5}a^2$, 2, $\frac{4}{3}$)\\ \hline
(3, 3, $a$)$\oplus$($\overline{3}$, 3, $-a$) & ($\frac{9}{5}a^2$, 8, 2)\\ \hline
(6, 1, $a$)$\oplus$($\overline{6}$, 1, $-a$) & ($\frac{6}{5}a^2$, 0, $\frac{10}{3}$)\\ \hline
(6, 2, $a$)$\oplus$($\overline{6}$, 2, $-a$) & ($\frac{12}{5}a^2$, 4, $\frac{20}{3}$)\\ \hline
(6, 3, $a$)$\oplus$($\overline{6}$, 3, $-a$) & ($\frac{18}{5}a^2$, 16, 10)\\ \hline
(8, 1, 0) & (0, 0, 2)\\ \hline
(8, 1, $a$)$\oplus$(8 1, $-a$) & ($\frac{8}{5}a^2$, 0, 4)\\ \hline
(8, 2, $a$)$\oplus$(8, 2, $-a$) & ($\frac{16}{5}a^2$, $\frac{16}{3}$, 8)\\ \hline
(8, 3, 0) & ($\frac{12}{5}a^2$, $\frac{32}{3}$, 6)\\ \hline
(8, 3, $a$)$\oplus$(8, 3, $-a$) & ($\frac{24}{5}a^2$, $\frac{64}{3}$, 12)\\ \hline
\end{tabular}
\caption{Contributions to $b_i$ from anomaly free fermions.
$U(1)_Y$ hypercharge "$a$" can take different values for different representations,
 and an electric charge is given by $Q_{\rm em} = I_3 + a/2$ with isospin $I_3$.
$b_1$ is given by $b_1=3/5 \times b_Y$.}
\label{bi_f}
\end{center}
\end{table}

\begin{table}[t]
\begin{center}
\begin{tabular}{|c|c|}\hline
Irreducible representation & Contribution to ($b_1$, $b_2$, $b_3$) \\
($SU(3)_C$, $SU(2)_L$, $U(1)_Y$) & by scalar particles\\
\hline \hline
(1, 1, $a$) & ($\frac{1}{20}a^2$, 0, 0)\\ \hline
(1, 2, $a$) & ($\frac{1}{10}a^2$, $\frac{1}{6}$, 0)\\ \hline
(1, 3, $a$) & ($\frac{3}{20}a^2$, $\frac{2}{3}$, 0)\\ \hline
(3, 1, $a$) & ($\frac{3}{20}a^2$, 0, $\frac{1}{6}$)\\ \hline
(3, 2, $a$) & ($\frac{3}{10}a^2$, $\frac{1}{2}$, $\frac{1}{3}$)\\ \hline
(3, 3, $a$) & ($\frac{9}{20}a^2$, 2, $\frac{1}{2}$)\\ \hline
(6, 1, $a$) & ($\frac{3}{10}a^2$, 0, $\frac{5}{6}$)\\ \hline
(6, 2, $a$) & ($\frac{3}{5}a^2$, 1, $\frac{5}{3}$)\\ \hline
(6, 3, $a$) & ($\frac{9}{10}a^2$, 4, $\frac{5}{2}$)\\ \hline
(8, 1, $a$) & ($\frac{2}{5}a^2$, 0, 1)\\ \hline
(8, 2, $a$) & ($\frac{4}{5}a^2$, $\frac{4}{3}$, 2)\\ \hline
(8, 3, $a$) & ($\frac{6}{5}a^2$, $\frac{16}{3}$, 3)\\ \hline
\end{tabular}
\caption{Contributions to $b_i$ by complex scalar particles.
$U(1)_Y$ hypercharge "$a$" can take different values for different representations,
 and an electric charge is given by $Q_{\rm em} = I_3 + a/2$ with isospin $I_3$.
Here, $b_1$ is normalized, i.e., $b_1=3/5 \times b_Y$.}
\label{bi_s}
\end{center}
\end{table}

Since the GCU is not realized in the SM,
 one has to extend the SM
 for the realization of GCU.
We will consider adding extra particles with the TeV scale mass to the SM
 without any additional gauge symmetry.
The extra particles with the TeV scale mass are motivated by avoiding the gauge hierarchy problem.
Once we fix extra particles,
 we can easily calculate the values of $b_i$ by using Table \ref{c2}.
However, we have to take care of gauge anomalies
 induced from extra fermions.
The simplest way to avoid the anomalies is to add extra fermions as a vector-like form.
Thus, in this paper, we will introduce the extra Weyl fermions as a vector-like form
 except for real representations such as (1,1,0), (1,3,0), (8,1,0), and (8,3,0),
 which do not yield any gauge anomaly.
Although the anomalies can be accidentally canceled as in the SM,
 we do not consider such cases.
Contributions of anomaly free fermions to $b_i$ are given in Table \ref{bi_f},
 which shows only small representations up to an adjoint representation, (8, 3, a).
In the same way,
 contributions from complex scalar particles to $b_i$ are given in Table \ref{bi_s}.
For real scalar particles, contribution to $b_i$ is half of the value in Table \ref{bi_s}
 because of $\eta$ (see Eq.\,(\ref{b})).

Next, we investigate conditions for the GCU.
The solution of Eq.\,(\ref{RGE}) are given by
\begin{eqnarray}
	\alpha_i^{-1}(M_{\rm GUT}) = \alpha_i^{-1}(M_*)-\frac{b_i}{2\pi}\ln\left(\frac{M_{\rm GUT}}{M_*}\right),
\label{alpha_GUT}
\end{eqnarray}
 where $M_*$ is the mass scale of extra particles
 and $M_{\rm GUT}$ is the GUT scale, in which the GCU can be realized.
The GCU conditions are given by
 $\alpha_i^{-1}(M_{\rm GUT})=\alpha_j^{-1}(M_{\rm GUT}) \equiv \alpha_{\rm GUT}^{-1}$ for $i,j=1$, 2, and 3.
Then, it can be written by
\begin{eqnarray}
	b'_i-b'_j = \frac{2\pi}{\ln\left(\frac{M_{\rm GUT}}{M_*}\right)}
				\left(\alpha_i^{-1}(M_*)-\alpha_j^{-1}(M_*)\right)-(b_i^{\rm SM}-b_j^{\rm SM})
\label{b_dif},
\end{eqnarray}
 where $b_i=b_i^{\rm SM}+b'_i$, and $b'_i$ are contributions of the extra particles.
Thus, once $M_*$ and $M_{\rm GUT}$ are fixed,
 one can see the required values of $b'_i$ for the realization of GCU.
In the following sections,
 we investigate possibilities for the realization of GCU at the Planck scale.

\section{General discussion for the GCU at the Planck scale} \label{sec:condition}
In this section,
 we investigate required values of $b'_i$ for the realization of GCU at the Planck scale.
Substituting $M_*=1$\,TeV and $M_{\rm Pl} \leq M_{\rm GUT} \leq \sqrt{8\pi}M_{\rm Pl}$ 
 ($M_{\rm Pl} = 2.4 \times 10^{18}\,{\rm GeV}$) into Eq.\,(\ref{b_dif}),
 we can find that the GCU can be realized when contributions of the extra particles satisfy
\begin{eqnarray}
	2.8 \lesssim &b'_3-b'_1& \lesssim 3.2 \label{b31},\\
	0.36 \lesssim &b'_3-b'_2& \lesssim 0.50 \label{b32},
\end{eqnarray}
 where the lower and upper bounds correspond to $M_{\rm GUT}=M_{\rm Pl}$ and $\sqrt{8\pi}M_{\rm Pl}$, respectively.
The RGEs and their boundary conditions in this analysis are given in Appendix.

In addition to these constraint,
 we impose the conditions of $\alpha_i^{-1}(M_{\rm GUT})>0$ to avoid the Landau pole
 (divergence of gauge couplings).
Then, these conditions lead
\begin{eqnarray}
	b'_i \lesssim \frac{2\pi}{\ln\left(\frac{M_{\rm GUT}}{M_*}\right)} \alpha_i^{-1}(M_*)-b_i^{\rm SM}.
\end{eqnarray}
As a result, $b'_i$ are limited to
\begin{eqnarray}
	b'_1 \lesssim 6.1\ (5.7),\qquad
	b'_2 \lesssim 8.6\ (8.4),\qquad
	b'_3 \lesssim 9.0\ (8.9),
 \label{b_max}
\end{eqnarray}
 where the values correspond to the $M_{\rm GUT}=M_{\rm Pl}$ ($\sqrt{8\pi}M_{\rm Pl}$) case.
Since all $b'_i$ are positive,
 gauge couplings become strong compared to those in the SM.
In Particular, extra fermions of large representations such as
 (6, 3, $a$)$\oplus$($\overline{6}$, 3, $-a$) in Table \ref{bi_f} cannot be added to the SM
 because both $b'_2$ and $b'_3$ are larger than the upper bound.
Similarly, extra particles with some large representations cannot also be added.
Thus, since we need not to consider higher representations than the adjoint representation,
 extra fermions in Table \ref{bi_f} are sufficient to investigate the realization of GCU.

\subsection{The GCU at the Planck scale by extra fermions}
When all extra particles are fermions,
 one can see that the smallest value of $b'_2$ and $b'_3$ are 2/3 from Table \ref{bi_f},
 and then $b'_3-b'_2 \propto 2/3$.
Thus, the cases of only extra fermions cannot satisfy Eq.\,(\ref{b32}),
 and unfortunately the GCU occurs at $M_{\rm GUT}\simeq 9.0\times 10^{16}$\,GeV or $7.8\times 10^{19}$\,GeV,
 for $b'_3-b'_2 = 0$ or 2/3, respectively.
This is the same result in Ref.~\cite{Giudice:2004tc}.
Note that, however, if we use two-loop RGEs and one-loop threshold corrections,
 the above results could be changed.
In fact, there exists ${\cal O}(1)$ uncertainty in values of gauge couplings at a high energy scale.
Thus, the GCU could be realized at the Planck scale even for $b'_3-b'_2 = 2/3$.
In addition,
 we can consider other possibility,
 in which extra fermions have different masses.
In Sec.\,\ref{sec:other}, we will show that the GCU at the Planck scale can be realized in this situation.

\subsection{The GCU at the Planck scale by extra fermions and scalars}

When extra particles include some scalars such as (1, 2, $a$),
 we can see that the smallest value of $b'_2$ and $b'_3$ are 1/6 from Table \ref{bi_s},
 and then $b'_3-b'_2 \propto 1/6$.
Then, there are two cases to satisfy Eq.\,(\ref{b32})
 in which the GCU is realized at the Planck scale as follows:
\begin{itemize}
\item
One is $b'_3-b'_2 = 1/3$, which corresponds to $M_{\rm GUT}\simeq M_{\rm Pl}$.\footnote{
In fact, since $b'_3-b'_2=1/3$ is a little below the lower bounds of Eq.\,(\ref{b32}),
$M_{\rm GUT}$ is also a little below $M_{\rm Pl}$ as $M_{\rm GUT}\simeq2.0\times10^{18}$\,GeV.}
In this case, $b'_1$ is determined by the lower bound of Eq.\,(\ref{b31}).
As a result, the GCU at $M_{\rm Pl}$ can be realized by extra particles satisfying
\begin{eqnarray}
	b'_3 = \frac{17}{6}+\frac{n}{6}\ (n=0,1,2,\cdots, {\rm and}\ 35),\qquad
	b'_2 = b'_3-\frac{1}{3},\qquad
	b'_1 \simeq b'_3-2.8,
\label{b_f}
\end{eqnarray}
 where the minimum value of $b'_3$ is determined to satisfy $b'_1 \geq 0$,
 and the largest value of $n$ is determined by Eq.\,(\ref{b_max}).

\item
Another is $b'_3-b'_2 = 1/2$,
 which corresponds to $M_{\rm GUT}\simeq \sqrt{8\pi}M_{\rm Pl}$
 because $b'_3-b'_2 = 1/2$ corresponds to upper bound of Eq.\,(\ref{b32}).
In this case, $b'_1$ is determined by the upper bound of Eq.\,(\ref{b31}).
Thus, the GCU at $\sqrt{8\pi}M_{\rm Pl}$ can be realized by extra particles satisfying
\begin{eqnarray}
	b'_3 = \frac{10}{3}+\frac{n}{6}\ (n=0,1,2,\cdots, {\rm and}\ 33),\qquad
	b'_2 = b'_3-\frac{1}{2},\qquad
	b'_1 \simeq b'_3-3.2,
\label{b_s}
\end{eqnarray}
 where the minimum value of $b'_3$ is determined to satisfy $b'_1 \geq 0$,
 and the largest value of $n$ is determined by the values in parentheses in Eq.\,(\ref{b_max}).
\end{itemize}

\begin{figure}[t]
  \begin{center}
          \includegraphics[clip, scale=1]{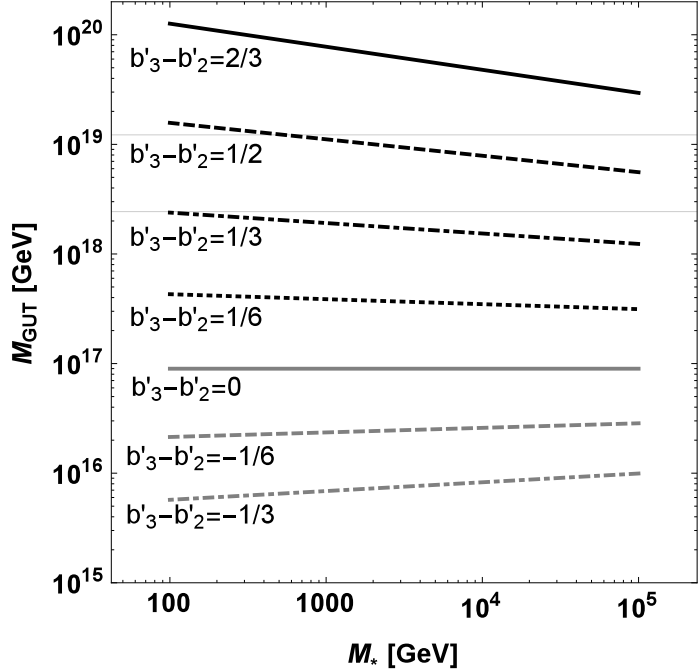}
  \end{center}
\caption{Relations between $M_*$ and $M_{\rm GUT}$ for fixed $b'_3-b'_2$.
These lines correspond to $b'_3-b'_2=$ 2/3, 1/2, $\cdots$, and $-1/3$, respectively.
Two horizontal lines represent the Planck scale,
 i.e. $M_{\rm Pl}=2.4\times10^{18}$\,GeV and $\sqrt{8\pi}M_{\rm Pl}=1.2\times10^{19}$\,GeV, respectively.}
\label{GUTscale}
\end{figure}

These results are understood by Eq.\,(\ref{b_dif}).
We show the relations between $M_*$ and $M_{\rm GUT}$ for fixed $b'_3-b'_2$ in Fig.\,\ref{GUTscale}.
The horizontal axis indicates $M_*$,
 and the vertical axis indicates $M_{\rm GUT}$, at which the GCU can be realized.
In the figure, each line corresponds to $b'_3-b'_2=$ 2/3, 1/2, $\cdots$, and $-1/3$.
We can see that $M_{\rm GUT}$ does not have a strong dependence of $M_*$
 once a value of $b'_3-b'_2$ is fixed.
It is worth noting that
 only $b'_3-b'_2=1/3$ or 1/2 can realize the GCU at the Planck scale,
 which are represented by two horizontal grid lines.
However, as mentioned in the previous subsection,
 if we use two-loop RGEs and one-loop threshold corrections,
 values of gauge couplings in a high energy scale could have ${\cal O}(1)$ uncertainty.
Thus, the GCU could be realized at the Planck scale even for $b'_3-b'_2 = 1/6$ and 2/3.

\section{Realization of the GCU at the Planck scale} \label{sec:realization}

\begin{table}[t]
\begin{center}
\begin{tabular}{|c|c|c|}\hline
 & Irreducible representation & Contribution to ($b_1$, $b_2$, $b_3$) \\
 & ($SU(3)_C$, $SU(2)_L$, $U(1)_Y$) & by fermions\\
\hline \hline
$Q\overline{Q}$ & (3, 2, $\frac{1}{3}$)$\oplus$($\overline{3}$, 2, $-\frac{1}{3}$) & ($\frac{2}{15}$, 2, $\frac{4}{3}$)\\ \hline
$U\overline{U}$ & (3, 1, $\frac{4}{3}$)$\oplus$($\overline{3}$, 1, $-\frac{4}{3}$) & ($\frac{16}{15}$, 0, $\frac{2}{3}$)\\ \hline
$D\overline{D}$ & (3, 1, $-\frac{2}{3}$)$\oplus$($\overline{3}$, 1, $\frac{2}{3}$) & ($\frac{4}{15}$, 0, $\frac{2}{3}$)\\ \hline
$L\overline{L}$ & (1, 2, $-1$)$\oplus$(1, 2, $1$) & ($\frac{2}{5}$, $\frac{2}{3}$, 0)\\ \hline
$E\overline{E}$ & (1, 1, $-2$)$\oplus$(1, 1, $2$) & ($\frac{4}{5}$, 0, 0)\\ \hline
$W$ & (1, 3, 0) & (0, $\frac{4}{3}$, 0)\\ \hline
\end{tabular}
\caption{Contributions to $b_i$ by the SM fermions (with vector-like partners) and adjoint fermions.}
\label{bi_SM}
\end{center}
\end{table}
\begin{table}[t]
\begin{center}
\begin{tabular}{|l||l|c|c|}\hline
Extra fermions & ($b'_1$, $b'_2$, $b'_3$) & $\alpha_{\rm GUT}^{-1}$ & $n$\\
\hline \hline
$Q\overline{Q} \times 1$ $\oplus$ $D\overline{D} \times 4$ $\oplus$ $W \times 1$ & ($\frac{6}{5}$, $\frac{10}{3}$, 4) & 28.0 & 7\\ \hline
$Q\overline{Q} \times 2$ $\oplus$ $D\overline{D} \times 3$ $\oplus$ $E\overline{E} \times 1$ & ($\frac{28}{15}$, 4, $\frac{14}{3}$) & 24.3 & 11\\ \hline
$Q\overline{Q} \times 2$ $\oplus$ $U\overline{U} \times 1$ $\oplus$ $D\overline{D} \times 2$ & ($\frac{28}{15}$, 4, $\frac{14}{3}$) & 24.3 & 11\\ \hline
$Q\overline{Q} \times 2$ $\oplus$ $D\overline{D} \times 4$ $\oplus$ $L\overline{L} \times 1$ $\oplus$ $E\overline{E} \times 1$ & ($\frac{38}{15}$, $\frac{14}{3}$, $\frac{16}{3}$) & 20.5 & 15\\ \hline
$Q\overline{Q} \times 2$ $\oplus$ $U\overline{U} \times 1$ $\oplus$ $D\overline{D} \times 3$ $\oplus$ $L\overline{L} \times 1$ & ($\frac{38}{15}$, $\frac{14}{3}$, $\frac{16}{3}$) & 20.5 & 15\\ \hline
$Q\overline{Q} \times 2$ $\oplus$ $U\overline{U} \times 2$ $\oplus$ $D\overline{D} \times 3$ $\oplus$ $W \times 1$ & ($\frac{16}{5}$, $\frac{16}{3}$, 6) & 16.8 & 19\\ \hline
$Q\overline{Q} \times 3$ $\oplus$ $U\overline{U} \times 2$ $\oplus$ $D\overline{D} \times 2$ $\oplus$ $E\overline{E} \times 1$ & ($\frac{58}{15}$, 6, $\frac{20}{3}$) & 13.1 & 23\\ \hline
$Q\overline{Q} \times 3$ $\oplus$ $U\overline{U} \times 3$ $\oplus$ $D\overline{D} \times 1$ & ($\frac{58}{15}$, 6, $\frac{20}{3}$) & 13.1 & 23\\ \hline
\end{tabular}
\caption{The leftmost column shows representations of extra fermions as ($SU(3)_C$, $SU(2)_L$, $U(1)_Y$).
With two $SU(2)_L$ doublets as (1, 2, 0), these extra fermions satisfy Eq.\,(\ref{b_f}).
In all cases, we take $M_*=1$\,TeV,
 and the GCU is realized at $M_{\rm Pl}$.
In the rightmost column, $n$ is given in Eq.\,(\ref{b_f}).
}
\label{combination}
\end{center}
\end{table}

According to the above discussions,
 we systematically investigate possibilities of the realization of GCU at the Planck scale,
 and find that a number of combinations of extra particles satisfy Eq.\,(\ref{b_f}) or (\ref{b_s}).
For simplicity,
 we consider
 representation of extra fermions are the same as the SM fermions (with vector-like partners)
 and an $SU(2)_L$ adjoint fermion
 as in Table \ref{bi_SM}.
Then, when we consider extra scalars are two $SU(2)_L$ doublets (1, 2, 0),
 the GCU can be realized at $M_{\rm Pl}$
 by extra fermions shown in Table \ref{combination}.\footnote{
Stable TeV-scale particles with fractional electric charge such as $SU(2)_L$ doublet scalar (1, 2, 0)
 might cause cosmological problems.
In order to avoid the problems,
 the reheating temperature after the inflation should be about 40 times lower than the particle masses~\cite{Chung:1998rq}.
In the case, the corresponding particles cannot be thermally produced in the universe.
Thus, since the reheating temperature should be larger than the QCD scale,
 we consider that it is ${\cal O}(10)\,{\rm GeV}$ in the case.
}
%
In all cases, masses of extra particles are 1\,TeV.
The values of the gauge couplings at $M_{\rm GUT}$ are calculated by Eq.\,(\ref{alpha_GUT}).
They are characterized by $n$ given in Eq.\,(\ref{b_f}),
 which is shown in the rightmost column.
The larger $n$ (equivalently $b_i$) becomes, the smaller $\alpha_{\rm GUT}^{-1}$ becomes.
We denote the pair of singlets (1, 1, $a$)$\oplus$(1, 1, $-a$) could be used
 for tuning the running of $g_1$ because it only affects $b_1$.
In addition, we did not list a complete gauge singlet fermion (1, 1, 0),
 which is usually considered as a right-handed neutrino,
 because this fermion does not affect the GCU.

\begin{figure}[t]
  \begin{center}
          \includegraphics[clip, scale=0.86]{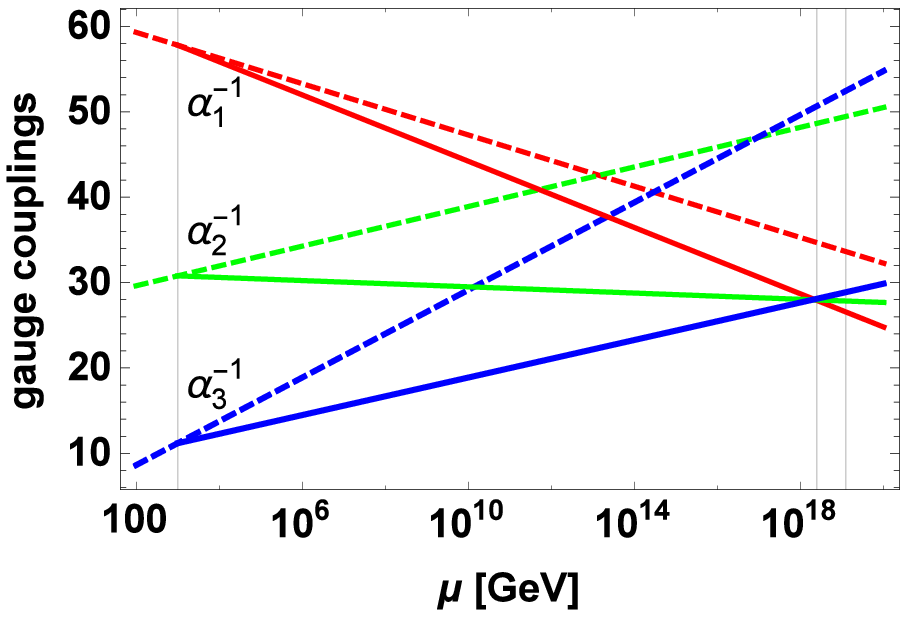}
          \includegraphics[clip, scale=0.86]{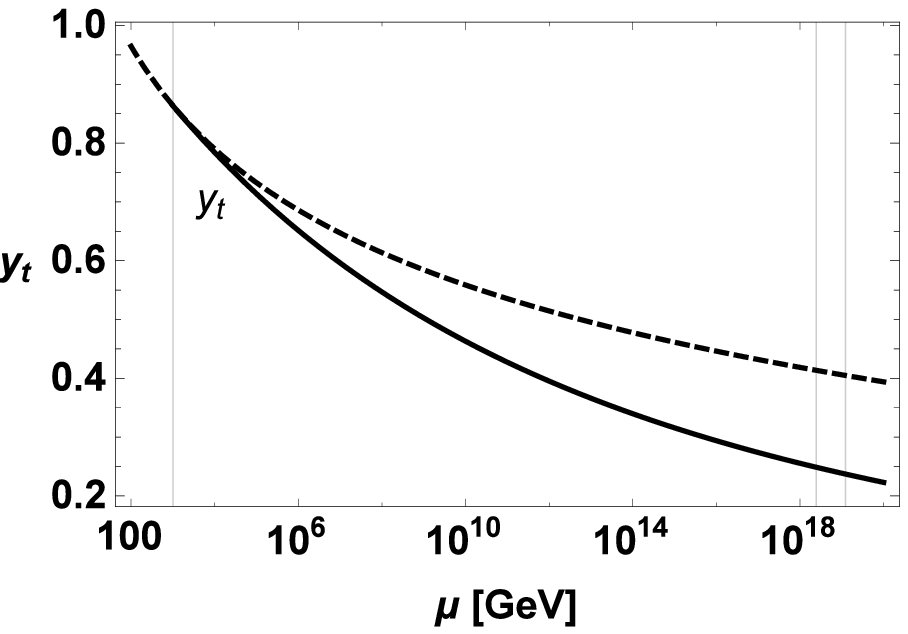}
  \end{center}
\caption{The runnings of gauge couplings (left figure),
 and top Yukawa coupling (right figure)
 in the extended SM
 where extra scalars are two $SU(2)_L$ doublets (1, 2, 0),
 and extra fermions are $Q\overline{Q} \times 1$ $\oplus$ $D\overline{D} \times 4$ $\oplus$ $W \times 1$,
 which correspond to the first one of Table \ref{combination}.
The solid and dashed lines correspond to the extended SM and the SM, respectively.
Three vertical lines represent $M_*$, $M_{\rm Pl}$, and $\sqrt{8\pi}M_{\rm Pl}$, respectively.}
\label{running}
\end{figure}

For a typical example, we consider the first one of Table \ref{combination}.
In Fig.\,\ref{running}
 we show the runnings of gauge and top Yukawa couplings in the extended SM model.
Here, we assume that coupling constants of extra particles to the SM particles are negligibly small,
 and thus introductions of the particles do not significantly change
 the runnings of top Yukawa and Higgs quartic couplings.
The solid and dashed lines correspond to the cases of the extended SM and the SM, respectively.
We can see that
 the GCU is realized at $M_{\rm Pl}$ as mentioned above.
In addition, the value of gauge couplings at $M_{\rm GUT}$ is
 $\alpha_{\rm GUT}^{-1}\simeq 28.0$ as in Table \ref{combination}.

From Fig.\,\ref{running},
 we can expect that the Higgs quartic coupling $\lambda$ is positive up to the Planck scale.
This reason is understood as follows.
In the extended SM, all gauge couplings are large compared to those in the SM
 because of $b_i\geq b_i^{\rm SM}$.
Then, $y_t$ becomes smaller due to the large gauge couplings (see Eq.\,(\ref{yt})).
Moreover, since $\beta_\lambda$ almost depends on quartic terms of $y_t$ and $g_i$,
 the smaller $y_t$ and the larger $g_i$ make $\beta_\lambda$ become larger (see Eq.\,(\ref{lambda})).
As a result, $\lambda$ becomes larger, and remains in positive value up to the Planck scale.
Even if mixing couplings between the Higgs boson and extra scalars are not negligible,
 contributions of the mixing couplings to $\beta_\lambda$ are positive
 as long as all of the mixing couplings are positive.
On the other hand,
 when extra fermions couple to the Higgs boson
 contributions of the couplings to $\beta_\lambda$ are negative.
Thus, in order to realize the vacuum stability,
 couplings between the Higgs boson and extra fermions should be small
 enough to satisfy $\lambda>0$.
Note that,
 when we neglect couplings between the SM particles and extra particles,
 we can see that $\lambda$ is positive up to the Planck scale.

\section{The GCU only with extra fermions} \label{sec:other}

Next, we consider other situations,
 in which extra fermions have different masses.
In the same way as before,
 we consider extra fermions within Table \ref{bi_SM}.
Moreover, their masses are taken as $0.5\,{\rm TeV} \leq M \leq 10\,{\rm TeV}$.
Actually, we take only lepton masses 0.5\,TeV,
 since lower bounds of vector-like lepton and quark masses are around 200\,GeV and 800\,GeV, respectively
 ~\cite{CMS:2012ra, Chatrchyan:2013uxa,Aad:2014efa}.
Unfortunately, we find that the GCU at $M_{\rm Pl}$ cannot be realized only by extra fermions.
In Table \ref{combination_f}, we show
 extra fermions which can realize the GCU around $\sqrt{8\pi}M_{\rm Pl}$.
Here, we relax the GCU condition as $\sqrt{8\pi}M_{\rm Pl} \lesssim M_{\rm GUT} \lesssim 2 \sqrt{8\pi}M_{\rm Pl}$
 because one-loop analyses always have ${\cal O}(1)$ ambiguity.
In the table, for example, "$W\times 1$ (0.5)" shows 
 one (1, 3, 0) fermions with a mass of 0.5\,TeV.
The reason why the GCU can be realized around $\sqrt{8\pi}M_{\rm Pl}$
 is understood by runnings of couplings as a following discussion.

\begin{figure}[t]
  \begin{center}
          \includegraphics[clip, scale=0.86]{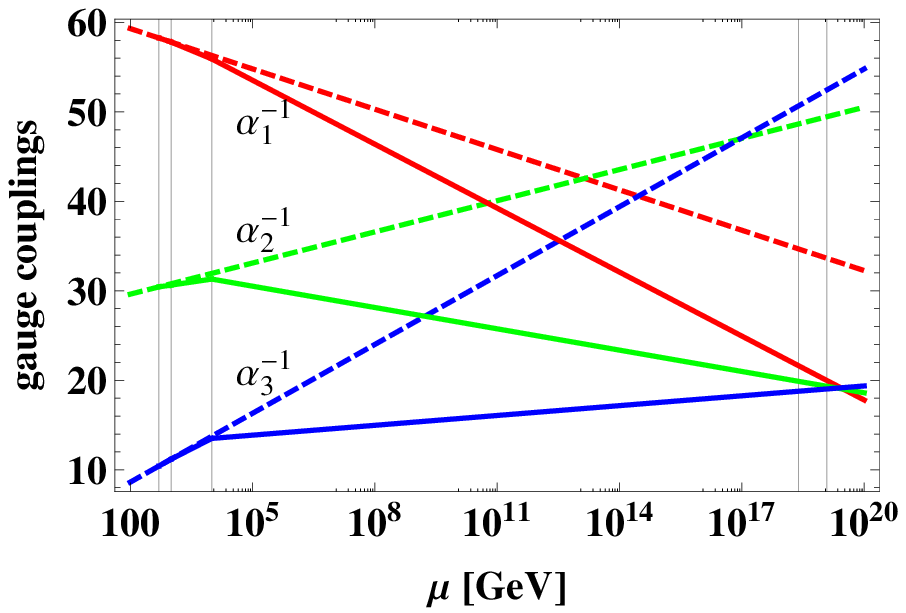}
          \includegraphics[clip, scale=0.86]{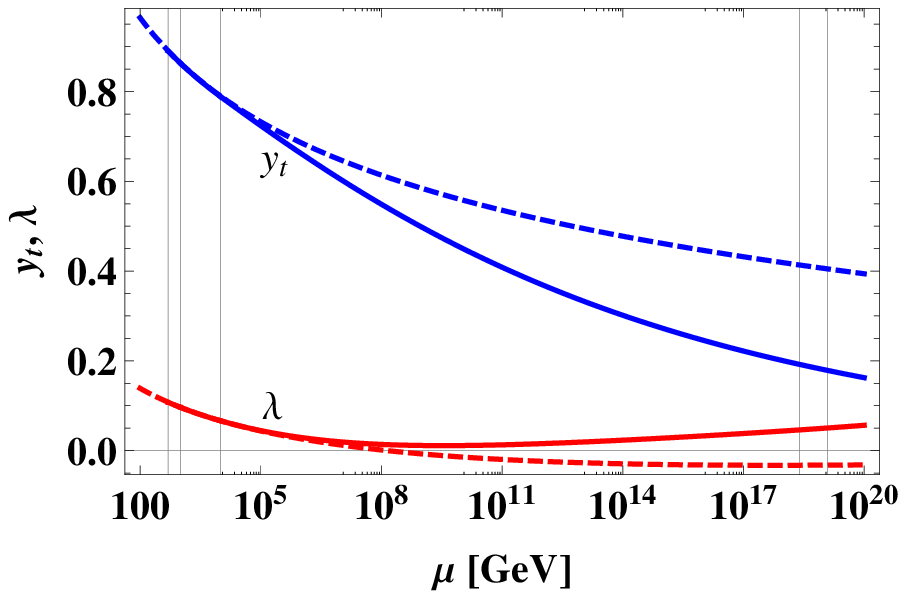}
  \end{center}
\caption{The runnings of gauge couplings (left figure),
 and top Yukawa coupling and Higgs quartic coupling (right figure)
 in the extended SM with extra fermions
 which correspond to the first one of Table \ref{combination_f}.
The solid and dashed lines correspond to the extended SM and the SM, respectively.
Three vertical lines represent
 0.5\,TeV, 3\,TeV, 10\,TeV, $M_{\rm Pl}$, and $\sqrt{8\pi}M_{\rm Pl}$, respectively.}
\label{running_f}
\end{figure}

In Fig.\,\ref{running_f}
 we show the runnings of gauge, top Yukawa, and Higgs quartic couplings in the extended SM model
 which correspond to the first one of Table \ref{combination_f}.
Here, we assume couplings between the Higgs doublet and extra fermions are negligibly small,
 and extra fermions do not significantly change running of top Yukawa and Higgs quartic couplings.
The solid and dashed lines correspond to the extended SM and the SM, respectively.
We can see that the GCU is realized around $\sqrt{8\pi}M_{\rm Pl}$.
When extra fermions have different masses,
 $\beta$-functions of gauge couplings change several times.
Then, our previous naive analyses are modified,
 and values of $M_{\rm GUT}$ shown in Fig.\,\ref{GUTscale} have ${\cal O}(1)$ uncertainty.
Thus, the GCU can be realized around $\sqrt{8\pi}M_{\rm Pl}$
 by extra fermions with $b'_3-b'_2=2/3$.
Note that, to realize the vacuum stability,
 couplings between the Higgs boson and extra fermions should be small as mentioned above.

\begin{table}[t]
\begin{center}
\begin{tabular}{|p{295pt}||l|c|}\hline
Extra fermions & ($b'_1$, $b'_2$, $b'_3$) & $\alpha_{\rm GUT}^{-1}$ \\
\hline \hline
$W\times 1$ (0.5) $\oplus$ $U\overline{U}\times 1$ (1) $\oplus$ $Q\overline{Q}\times 2$ (10) $\oplus$ $D\overline{D}\times 4$ (10) & ($\frac{12}{5}$, $\frac{16}{3}$, 6) & 19.1 \\ \hline
$E\overline{E}\times 2$ (0.5) $\oplus$ $Q\overline{Q}\times 2$ (2) $\oplus$ $Q\overline{Q}\times 2$ (10) $\oplus$ $D\overline{D}\times 4$ (10) & ($\frac{46}{15}$, 6, $\frac{20}{3}$) & 14.9 \\ \hline
$L\overline{L}\times 1$ (0.5) $\oplus$ $E\overline{E}\times 1$ (0.5) $\oplus$ $Q\overline{Q}\times 1$ (1) $\oplus$ $U\overline{U}\times 1$ (1) $\oplus$ $Q\overline{Q}\times 2$ (10) $\oplus$ $D\overline{D}\times 4$ (10) & ($\frac{56}{15}$, $\frac{20}{3}$, $\frac{22}{3}$) & 11.1 \\ \hline
$E\overline{E}\times 1$ (0.5) $\oplus$ $W\times 1$ (0.5) $\oplus$ $U\overline{U}\times 2$ (4) $\oplus$ $Q\overline{Q}\times 3$ (10) $\oplus$ $D\overline{D}\times 4$ (10) & ($\frac{22}{5}$, $\frac{22}{3}$, 8) & 7.95 \\ \hline
\end{tabular}
\caption{Examples of combinations of extra fermions
 which realize the GCU around $\sqrt{8\pi}M_{\rm Pl}$.
In the leftmost column, the characters show extra fermions as in Table \ref{bi_SM},
 and the values in bracket show the fermion masses with a unit of TeV.
}
\label{combination_f}
\end{center}
\end{table}

Finally, we mention the GCU at the string scale ($M_{\rm GUT}=\Lambda_s \approx 5.27 \times 10^{17}$\,GeV).
Figure \ref{GUTscale} shows that
 the GCU at the string scale could be realized by $b'_3-b'_2=0$, 1/6, and 1/3.
The ${\cal O}(1)$ difference could come from two-loop RGEs and one-loop threshold corrections.
On the other hand, another possibility is discussed in Ref.~\cite{Barger:2006fm}.
In this paper, the authors consider several possible string-GUT models.
Then, the GCU condition is given by
\begin{eqnarray}
	\alpha_{\rm string} = \frac{2 G_N}{\alpha'} = k_i \alpha_i,
\end{eqnarray}
 where $G_N$ and $\alpha'$ are the gravitational constant and the Regge slope, respectively.
The factor $k_i$ ($i=Y$, 2, and 3) is the so-called Ka$\breve{c}$-Moody levels,
 and the values are different for the considering GUT models~\cite{PerezLorenzana:1999tf}.
Particularly, $k_2$ and $k_3$ should be positive integer,
 and we take Ka$\breve{c}$-Moody levels as ($k_Y$, $k_2$, $k_3$) = (5/3, 1, 1),
 which are given in GUT models such as $SU(5)$ and $SO(10)$.
However, for $k_2\neq1$ and/or $k_3 \neq 1$, the GCU conditions of our analyses are changed.
When the new physic scale is $M_*=1$\,TeV,
 the GCU at $\Lambda_s$ can be realized by
\begin{eqnarray}
	\frac{b'_3}{2}-b'_2 \simeq -4.34 \approx -\frac{13}{3},\qquad
	\frac{b'_3}{2}-\frac{3}{13}b'_1 \simeq 1.99
\end{eqnarray}
 for ($k_Y$, $k_2$, $k_3$) = (13/3, 1, 2),
 which is given in the GUT model as $SU(5)\times SU(5)$ and $SO(10)\times SO(10)$.
In the same way, the GCU at $\Lambda_s$ can be realized by
\begin{eqnarray}
	b'_3-\frac{b'_2}{2} \simeq 4.64 \approx \frac{14}{3},\qquad
	b'_3-\frac{3}{2}b'_1 \simeq -7.46
\end{eqnarray}
 for ($k_Y$, $k_2$, $k_3$) = (2/3, 2, 1),
 which is given in the GUT model as $E_7$.
Both conditions can be satisfied only by extra fermions
 due to $b'_3/2-b'_2 \propto 1/3$ and $b'_3-b'_2/2 \propto 1/3$.
Thus, in some string-GUT models,
 the GCU at $\Lambda_s$ can be realized only by extra fermions.

\section{Summary and discussion} \label{sec:summary}
We have explored possibilities of GCU at the Planck scale in the extended SM
 which includes extra particles around the TeV scale.
To avoid the gauge anomaly, extra fermions are constrained as vector-like and adjoint representations.
When all extra particles are fermions and their masses are the same,
 the GCU at the Planck scale cannot be realized (up to one-loop level).
On the other hand, when extra particles include some scalar particles
 there are two cases which realize the GCU at the Planck scale.
The conditions of the GCU at $M_{\rm Pl}$ and $\sqrt{8\pi}M_{\rm Pl}$ are given
 by Eqs.\,(\ref{b_f}) and (\ref{b_s}), respectively.
Then, we have found that there are a number of combinations which satisfy these equations.
For examples, when extra scalars are two $SU(2)_L$ doublets as (1, 2, 0),
 the GCU at $M_{\rm Pl}$ are realized by extra fermions given in the leftmost column of Table \ref{combination}.

Moreover, we have considered other situations,
 in which extra fermions have different masses.
In this case, extra fermions can realize the GCU around $\sqrt{8\pi}M_{\rm Pl}$
 as in Table \ref{combination_f}.
Since $\beta$-functions of gauge couplings change several times by extra fermions with different masses,
 our previous naive analyses are modified,
 and the GCU can be realized around $\sqrt{8\pi}M_{\rm Pl}$.
Note that, if we use the two-loop RGEs and one-loop threshold corrections,
 these results could change,
 and other possibilities could exist.

If there are no intermediate scales between the TeV scale and the GCU scale,
 and quantum corrections to the Higgs mass term are completely vanishing at the GCU scale
 due to a UV-complete theory,
 the Higgs mass receives quantum corrections only from TeV scale particles.
In this paper, we have assumed that the GCU scale is the Planck scale,
 and the Higgs mass term are vanishing at the scale.
More detailed discussion has been done in the introduction and Ref.\,\cite{Iso:2012jn}.
When the GCU at the Planck scale is realized,
 gauge couplings become larger compared to the SM case.
Then, top Yukawa and Higgs quartic couplings become smaller and larger, respectively.
As a result, the vacuum can be stable up to the Planck scale.

Finally, we mention the proton lifetime in a GUT model.
Although we do not discuss any specific GUT model,
 the proton lifetime should be long enough to avoid the experimental lower bound.
The proton lifetime is usually given by
\begin{eqnarray}
	\tau_{\rm proton} \sim \left( \alpha_i^{-1}(M_{\rm GUT}) \right)^2 \frac{M_{\rm GUT}^4}{m_{\rm proton}^5}.
\end{eqnarray}
This is derived from a four-fermion approximation for the decay channel
 $p \rightarrow e^+ + \pi^0$.
For $M_{\rm GUT}\simeq M_{\rm Pl}$,
 we obtain $\tau_{\rm proton} \sim \left( \alpha_i^{-1}(M_{\rm GUT}) \right)^2\times 10^{42}$ yrs.
Since $\alpha_i^{-1}(M_{\rm GUT})$ is larger than 1 (see Table \ref{combination}),
 the proton lifetime is much longer than the experimental lower bound.

\subsection*{\centering Acknowledgment} \label{Acknowledgement}
We thank H. Sugawara and N. Okada for helpful and fruitful discussions which
leaded a motivation of this research.
We also thank S. Iso and T. Yamashita for helpful discussions and valuable comments.
This work is partially supported by Scientific Grant
 by the Ministry of Education, Culture, Sports, Science and Technology,
 No. 24540272.
The works of R.T. and Y.Y. are supported
 by Research Fellowships of the Japan Society for the Promotion of Science for Young Scientists
 (Grants No. 24$\cdot$801 (R.T.) and No. 26$\cdot$2428 (Y.Y.)).

\section*{Appendix}
\appendix
\section*{{\boldmath $\beta$-functions} in the SM} \label{app:RGE}
The RGE of coupling $x$ is given by $dx/d\ln \mu = \beta_x$, in which $\mu$ is a renormalization scale.
The $\beta$-functions in the SM are given by
\begin{eqnarray}
	&&\beta_{g_1} = \frac{g_1^3}{(4\pi)^2} \left[ \frac{41}{10} \right],\\
	&&\beta_{g_2} = \frac{g_2^3}{(4\pi)^2} \left[ -\frac{19}{6} \right],\\
	&&\beta_{g_3} = \frac{g_3^3}{(4\pi)^2} \left[ -7 \right],\\
	&&\beta_{y_t} = \frac{y_t}{(4\pi)^2} \left[ \frac{9}{2}y_t^2 - \frac{17}{20}g_1^2 - \frac{9}{4}g_2^2
		- 8 g_3^2 \right], \label{yt}\\
	&&\beta_{\lambda}  = \frac{1}{(4\pi)^2} \left[ \lambda \left( 24 \lambda + 12 y_t^2 - \frac{9}{5}g_1^2
		- 9 g_2^2 \right) - 6 y_t^4 + \frac{27}{200}g_1^4 + \frac{9}{8}g_2^4 + \frac{9}{20}g_1^2 g_2^2 \right] \label{lambda},
\end{eqnarray}
 up to one-loop level~\cite{Buttazzo:2013uya}.
We have only included the top quark Yukawa coupling,
 and omitted the other Yukawa couplings,
 since they do not contribute significantly to the Higgs quartic coupling and gauge couplings.

To solve the RGEs, we take the following boundary conditions \cite{Buttazzo:2013uya}:
\begin{eqnarray}
	&&g_Y(M_t) = 0.35761 + 0.00011 \left( \frac{M_t}{{\rm GeV}} - 173.10 \right), \qquad g_1 = \sqrt{\frac{5}{3}}g_Y, \label{g1_mt}\\
	&&g_2(M_t) = 0.64822 + 0.00004 \left( \frac{M_t}{{\rm GeV}} - 173.10 \right),\\
	&&g_3(M_t) = 1.1666 - 0.00046 \left( \frac{M_t}{{\rm GeV}} - 173.10 \right) + 0.00314 \left( \frac{\alpha_3(M_Z) - 0.1184}{0.0007} \right),\\
	&&y_t(M_t) = 0.93558 + 0.00550 \left( \frac{M_t}{{\rm GeV}} - 173.10 \right) -0.00042 \left( \frac{\alpha_3(M_Z) - 0.1184}{0.0007} \right), \label{yt_mt}\\
	&&\lambda(M_t) = 0.12711 - 0.00004 \left( \frac{M_t}{{\rm GeV}} - 173.10 \right) + 0.00206 \left( \frac{M_h}{{\rm GeV}} - 125.66 \right), \label{lambda_mt} \\
	&&\alpha_3(M_Z) = 0.1184 \pm 0.0007,
\end{eqnarray}
 where $M_t$ and $M_h$ are the pole masses of top quark and Higgs boson, respectively.
In this paper, we have used $M_t = 173.3$\,GeV, $M_h=125.7$\,GeV and $\alpha_3(M_Z)=0.1184$.



\end{document}